\documentclass[aps,pre,reprint,superscriptaddress]{revtex4-1}

\usepackage[dvips]{graphicx}
\usepackage{bm}
\usepackage{color}
\usepackage[version=3]{mhchem}

\begin{document}


\title{Thermonuclear fusion triggered by collapsing standing whistler
  waves in magnetized overdense plasmas}%


\author{Takayoshi Sano}
\email{sano@ile.osaka-u.ac.jp}
\affiliation{Institute of Laser Engineering, Osaka University, Suita,
  Osaka 565-0871, Japan} 


\author{Shinsuke Fujioka}
\affiliation{Institute of Laser Engineering, Osaka University, Suita,
  Osaka 565-0871, Japan} 

\author{Yoshitaka Mori}
\affiliation{The Graduate School for the Creation of New Photonics
  Industries, Hamamatsu, Shizuoka 431-1202, Japan} 

\author{Kunioki Mima}
\affiliation{Institute of Laser Engineering, Osaka University, Suita,
  Osaka 565-0871, Japan} 
\affiliation{The Graduate School for the Creation of New Photonics
  Industries, Hamamatsu, Shizuoka 431-1202, Japan} 

\author{Yasuhiko Sentoku}
\affiliation{Institute of Laser Engineering, Osaka University, Suita,
  Osaka 565-0871, Japan} 


\date{Jan 7, 2020; accepted for publication in Physical Review E}

\begin{abstract}
Thermal fusion plasmas initiated by standing whistler waves are
investigated numerically by two- and one-dimensional 
Particle-in-Cell simulations.
When a standing whistler wave collapses due to the wave breaking of ion
plasma waves, the energy of the electromagnetic waves transfers
directly to the ion kinetic energy.   
Here we find that the ion heating by the standing whistler wave is
operational even in multi-dimensional simulations of multi-ion species
targets, such as deuterium-tritium (DT) ices and solid ammonia borane
(H$_6$BN). 
The energy conversion efficiency to ions becomes as high as 15\% of
the injected laser energy, which depends significantly on the
target thickness and laser pulse duration. 
The ion temperature could reach a few tens of keV or much higher if
appropriate laser-plasma conditions are selected. 
DT fusion plasmas generated by this method must be useful as efficient
neutron sources. 
Our numerical simulations suggest that the neutron generation
efficiency exceeds 10$^9$ n/J per steradian, which is beyond the
current achievements of the state-of-the-art laser experiments.
The standing whistler wave heating would expand the experimental
possibility for an 
alternative ignition design of magnetically confined laser fusion, and
also for more difficult fusion reactions including the aneutronic
proton-boron reaction.  
\end{abstract}


\maketitle


\section{Introduction}

The ultimate goal of fusion science is to heat ions to the high enough
temperature to promote fusion reactions.
In the laser-driven inertial confinement fusion (ICF) scheme, the
laser energy should be converted to the ion energy of fuels as much as possible
through various laser-plasma interactions \cite{lindl95,atzeni04}. 
However, there is a fundamental problem that electrons take away most
of the laser energy at the first step of the interaction.
The energy conversion from the electrons to ions occurs slowly
via collisional processes, and the efficiency is not high in general. 
Therefore the development of direct energy transfer from electromagnetic waves to ions has a valuable meaning that overcomes
the essential difficulty.
In the central ignition scheme, the role of the implosion is
responsible for the compression and heating of the fuel at the same time.
On the other hand, the fast ignition scheme is based on the idea of
``division of labor'', in which two kinds of lasers are specialized
separately to the increase of the density and temperature
\cite{basov92,tabak94,kodama01}. 
The problem in the fast ignition scheme is how to increase the ion temperature of the imploded core plasmas \cite{fujioka16,sakata18,matsuo19}.

The motivation of our work is to explore an efficient mechanism of
the energy transfer from the lasers to overdense ions directly.
A possible mechanism of ion heating proposed recently is caused by
collapsing standing whistler waves \cite{sano19}.
The advantage suitable for the fast ignition scheme is the high efficiency
of the energy transfer to ions in an ultrafast timescale of the order of
the laser period.
The whistler wave is an electromagnetic wave traveling along a
magnetic field line. 
The tangential electric field is right-hand circularly polarized (CP)
to the direction of the external magnetic field.
Thanks to the no-cutoff nature of the whistler wave, the laser energy is
delivered to the inside of overdense plasmas, which enables
the direct interaction of the electromagnetic waves and ions there.
In overdense plasmas, electrons are treated as a fluid in a timescale
of the laser period, which is one of the essential assumptions for
the ion heating by standing whistler waves.
If the electron density is subcritical, the energy of the standing
wave is transferred to the electrons predominantly through the
stochastic acceleration \cite{hsu79,doveil81,tran82}. 

The hardest requirement of the standing whistler wave heating is a
strong magnetic field 
larger than the critical strength $B_c \approx 10 (\lambda_0 / 1 \;
\mu{\rm m})^{-1}$ kT, where $\lambda_0$ is the laser wavelength
{\color{black} in the vacuum}.
The achievement of strong magnetic fields of kilo-Tesla order in laser
experiments has been reported by several groups
\cite{yoneda12,fujioka13,korneev15,goyon17,santos18}. 
If such the laser-induced magnetic field is compressed by the implosion, the field strength might be enhanced by an order of magnitude and reach the critical value $B_c$.
Although the current status of the maximum strength is not yet
sufficient for the propagation of the whistler 
waves in overdense plasmas, it will become plausible shortly
to excite relativistic whistler waves from high-intensity lasers under
a supercritical field condition $B_{\rm ext} > B_c$.
We believe the theoretical exploration of tens of kT or more is
worth to be attempted because it may bring a drastic change in the
qualitative consideration of ion heating by lasers. 
Furthermore, the critical field strength is inversely proportional to
the laser wavelength so that this mechanism could happen with an
appropriate field strength in other situations in space and laboratory
plasmas \cite{stenzel99,stenzel16,stacey05}.

The standing whistler wave heating is suitable for the generation of
thermal neutrons because of the high efficiency of the energy
conversion from the heating laser to ions.
Even though the energy gain is not adequate for the fusion energy
application, the laser-driven neutron source has a wide range of
applications in various fields such as material science, biology,
medicine, security, and industrial engineering \cite{alvarez14}. 
There are two popular methods to generate neutrons in laser
experiments.
The first method is spallation schemes which use high-energy beams of
protons or deuterons accelerated by short-pulse lasers from a thin foil target
\cite{norreys98,lancaster04,higginson11,roth13,zulick13,kar16,kleinschmidt18}.
Directive neutron sources are generated when the deuteron beams hit
the secondary target of beryllium, for instance \cite{roth13,kleinschmidt18}.
The neutron yield in the laser direction is $10^9$ neutrons per steradian
that originated from the laser energy of less than 100 J \cite{kleinschmidt18}.
The second one is the thermal neutron generation through the implosion
by high-power lasers \cite{yamanaka86,chen90,doppner15,regan16,lepape18}.
The record result of the total number of neutrons is $10^{16}$ by
using 1 MJ energy of the National Ignition Facility \cite{lepape18}.
The total neutron number in the implosion method is normally more
substantial than those in the beam fusion.
However, the implosion method is not suitable for the compact source
because it requires at least the kJ-class lasers for the implosion.
Our method utilized by the whistler waves could be 
regarded as a new approach to generate
the thermal fusion plasmas even with Joule-class compact heating lasers.  

In this paper, we extended the numerical simulations of the standing
whistler wave heating to multi-dimensional and multi-ion species targets such as the
deuterium-tritium (DT) mixture to evaluate the neutron yield. 
In the previous paper \cite{sano19}, we have built the theoretical
framework of the standing whistler wave heating, which has
been confirmed by one-dimensional (1D) Particle-in-Cell (PIC)
simulations for pure hydrogen plasmas. 
The plan of this paper is as follows.  
Numerical setup and method of 2D PIC simulations
{\color{black} (2D in space and 3D in velocity)}
are described in Section~\ref{sec2}. 
In Section~\ref{sec3}, the mechanism and characteristics of the
standing whistler wave heating is briefly revisited based on the
theoretical model in our previous paper \cite{sano19}.
Multi-dimensional and multi-ion spices effects are examined in
Section~\ref{sec4}.
The careful comparisons between 2D and 1D outcomes are carried out.
We derive the optimized conditions for the highest conversion efficiency
of the laser energy to ions by a systematic survey of various 1D
PIC simulations.
Neutron yield expected from the DT target heated by standing whistler
waves is also evaluated numerically from long-term simulations. 
In Section~\ref{sec5}, we discuss an alternative design of the fast
ignition scheme and a possibility of the aneutronic proton-boron
fusion reaction.  
Finally, our conclusions are summarized in Section~\ref{sec6}.

\section{Numerical Setup and Method \label{sec2}}

As for the initial setup to simulate the standing whistler wave
heating, we investigate a simple configuration in which two CP lasers irradiate a thin overdense target from both sides.
We perform numerical simulations both in the 2D and 1D
Cartesian coordinate.
The $x$ axis is always defined by the direction of a uniform external magnetic field with the strength of $B_{\rm ext}$.
The counter lasers are injected parallel to the magnetic field line in
the $\pm x$ direction.
The physical quantities are initially uniform in
the $y$ direction.
The laser wavelength is set to be $\lambda_0 = 1$ $\mu$m in all
the calculations in this paper.   

The thickness of the target foil is $L_x$, and the material is a solid DT mixture.
The origin of $x$ is defined at the center of the target, so that
the DT foil is located at $|x| \le L_x / 2$.
For simplicity, a fully ionized plasma with zero temperature is assumed at
the beginning.
The densities of the deuterons and tritons are constant in the target,
$n_D = n_T = n_{e0} / 2$, where $n_{e0}$ is the initial electron density.
The outside of the target is set to be the vacuum region. 

The CP lasers with the frequency $\omega_0$ are injected from both
sides of the $x$ boundaries of the computational domain. 
The polarization of the two lasers is right-hand circular toward
the magnetic field direction.
The normalized vector potential $a_0 = e E_0 / (m_e c \omega_0)$
characterizes the amplitude of the laser electric field $E_0$ in the
vacuum, where $e$ is the elementary charge, $m_e$ is the electron
mass, and $c$ is the speed of light. 
The intensity of a CP light is expressed as $I_0 = \epsilon_0 c
E_0^2$, where $\epsilon_0$ is the vacuum permittivity.
The envelope of the laser field is a flat-top shape where 
the pulse duration is $\tau_0$ and the rise time is of the
order of the laser period.
In our problem setup, the laser lights coming from the boundaries
propagate in the vacuum until they hit the target surface.
The refractive index of the whistler wave $N$, which is a function of the plasma density $n_{e0}$ and magnetic field strength $B_{\rm ext}$, determines the transmittance and reflectivity at the target surface.

Numerical simulations on the laser-plasma interaction under an external
magnetic field are calculated by a PIC scheme, PICLS \cite{sentoku08}, in which the Coulomb collision term is solved by a fully relativistic energy-conserving binary collision model.  
In the $x$ direction, the computational domain size $D_x$ is broader than the target thickness.
The escape boundary conditions for the electromagnetic waves and
plasma particles are applied at both of the $x$ boundaries.
The domain size in the additional spatial direction $y$ is $D_y$. 
The periodic boundary conditions are adopted in the $y$ direction because the foil length is intended to be much longer than the domain
size. 
The laser spot size is assumed implicitly to be larger than $D_y$ so that
the spatial profile of the laser injection at the $x$ boundaries is
independent of $y$. 

The spatial and time resolutions in the 2D (1D) runs are $\Delta x = c
\Delta t = \lambda_0 / 300$ ($\lambda_0 / 1000$) and the particle
number is 60 (200) per each grid initially, unless otherwise mentioned.
The third order interpolation algorithm is applied to suppress the
numerical heating, and also the time resolution $\Delta t$ should
be sufficiently smaller than the electron gyration time as well as the
plasma oscillation time to avoid the unphysical numerical heating in
strongly magnetized plasmas.  
The wavelength of whistler waves in the dense targets
is shorter than that in the vacuum $\lambda_0$. 
In order to capture the evolution of the standing whistler waves
correctly, the whistler wavelength $\lambda_w$ should be resolved by at least a hundred of grids. 
These criteria are satisfied in all simulations shown in this paper.
It has been tested by the convergence check that the qualitative
conclusions are unaffected by the numerical resolution. 

\section{Theoretical Background \label{sec3}} 

Before moving to the numerical results, theoretical aspects of the ion heating by collapsing standing CP waves are summarized in this section.     

The ion heating occurs when standing waves of CP electromagnetic waves are formed inside of overdense plasmas where the electron density is larger than the critical value of $n_c = \epsilon_0 m_e \omega_0^2 / e^2$.
For the propagation of electromagnetic waves in overdense materials,
the existence of an external magnetic field is an inevitable requirement. 
The CP waves propagating along a magnetic field line are called the R- or L-waves depending on the direction of polarization against the orientation of the magnetic field.
When the external magnetic field $B_{\rm ext}$ exceeds the critical
strength $B_c = m_e \omega_0 / e$, the whistler branch of the right-hand CP lights appears at the lower frequency in the dispersion relation.
The whistler wave can propagate at any density of plasmas because of no cutoff density, and then have an opportunity to interact directly with overdense plasmas. 
The standing whistler wave appears naturally in head-on collisions
of two counter-propagating whistler waves. 
The left-hand CP lights, on the other hand, have the L-cutoff density
defined by $n_L = n_c (1 + B_{\rm ext}/B_c)$.
Although the L-waves can also propagate in overdense situations under the supercritical magnetic field $B_{\rm ext} > B_c$, we concentrate our
studies in this paper to the cases of the whistler waves.
It is because the higher density and relatively weaker field strength are preferable plasma conditions along with our motivation.

Let us consider the counter configuration of whistler waves.
The electromotive force in the longitudinal direction $( \bm{v \times
  B})_x$ is always zero everywhere in a single whistler wave, where
$\bm{v}$ and $\bm{B}$ are the tangential component of the electron
quiver velocity and magnetic field of the whistler eigenmode. 
However, it becomes finite when the two waves are overlapped, that is,
in the standing wave.
The amplitude of the longitudinal force is constant with time and
sinusoidal in the $x$ direction with a period of half of the
whistler wavelength.  
Driven by the constant force, the
standing whistler waves collapse immediately through the nonlinear
wave breaking.
A large fraction of the wave energy is converted directly to the ions during the rapid collapse phase.

\begin{figure}
\includegraphics[scale=0.6,clip]{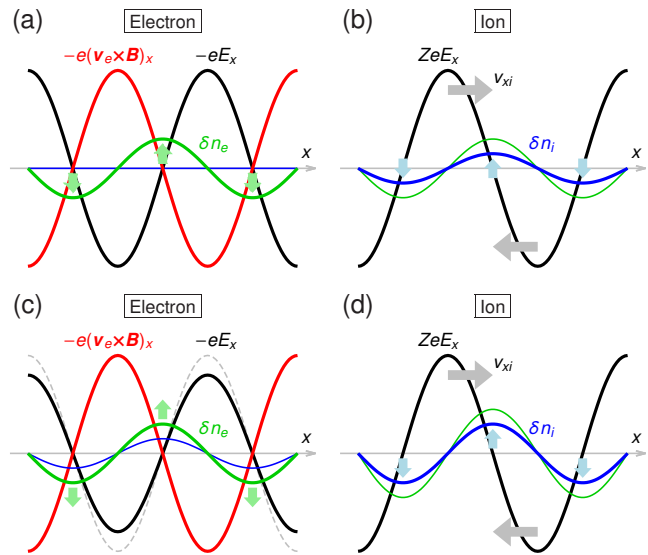}%
\caption{
Schematic sketches of the force and density distributions of
the electrons and ions during the collapse of a standing whistler
wave.
The horizontal axis is the spatial direction of $x$.
(a) The density distribution of the electron fluid (green curve) is modified quickly in response to the appearance of the electromotive force (red curve) associated with the standing whistler wave.
The force balance is established by the longitudinal electric field
(black curve).
(b) The longitudinal electric field accelerates the ions, which move toward the antinodes of the standing wave.
The blue curve shows the ion-density fluctuation.
(c) The electric field is sustained by the redistribution of the
electron fluid as long as the standing wave exists.
(d) The ions continue to be accelerated effectively by the constant
electric field until the wave breaking takes place.
\label{fig1}}
\end{figure}

The schematic drawings of the evolution of a standing whistler wave are shown in Fig.~\ref{fig1}.
In overdense plasmas where the electron plasma frequency is much higher than the laser frequency, the electrons behave like a fluid in the timescale of the laser period.
The appearance of the electromotive force initiates the redistribution of the electron density.
The electrons move toward the antinodes of the standing wave, and then the longitudinal electric field is generated in balance with the
electromotive force $E_x = - ( \bm{v \times B})_x$ [Fig.~\ref{fig1}(a)].  
On the other hand, the electromotive force is ineffective in the ion
equation of motion. 
The ions are accelerated by the longitudinal electric field and start to move toward the antinodes [Fig.~\ref{fig1}(b)].  
Then the ion density fluctuation reduces the amplitude of the electric field $E_x$.
However, the electromotive force is unchanged as long as the standing
wave exists.
The condensation of the electron density proceeds furthermore to balance
the longitudinal force in the electron fluid.
As a result, the longitudinal electric field is sustained at the
constant amplitude [Fig.~\ref{fig1}(c)].  
This positive feedback in terms of the ion acceleration is ensured
in the standing whistler wave until the ion acceleration is disturbed
by the wave breaking  [Fig.~\ref{fig1}(d)]. 

The substantial enhancement of the density fluctuation leads to the collapse of the standing wave.
After the ion-wave breaking, counter ion beams coexist at many locations.
Finally, most of the accelerated ion energy is thermalized through the
kinetic two-stream instabilities.
Because of the periodic structure of the longitudinal electric field
with a tiny scale of the whistler wavelength, 
the excellent advantage in a series of these processes is to contain
the efficient mechanism of ion acceleration and thermalization
simultaneously.

For the interpretation of the simulation results, it would be useful to
enumerate some quantitative evaluations related to the ion-heating mechanism by the standing whistler waves \cite{sano19}.
In the following discussion, the variables with tilde on top stand for the normalized non-dimensional quantities.
The density and magnetic field strength are given in terms of the
critical values $n_c$ and $B_c$, i.e., $\widetilde{n}_{\ast} \equiv
n_{\ast} / n_c$ and $\widetilde{B}_{\ast} \equiv B_{\ast} / B_c$.
The length, time, and velocity are normalized by $\lambda_0$,
$\omega_0^{-1}$, and $c$, respectively.
The electric field is given in the non-dimensional form of
$a_{\ast} \equiv e E_{\ast} / (m_e c \omega_0)$. 

The amplitude of the electromotive potential is roughly estimated by $\psi \sim e v_w^{\pm} B_w^{\mp} / k_w$, where $v_w^{\pm}$ and
$B_w^{\pm}$ are the amplitude of velocities and magnetic fields of a
pair of counter whistler waves with the wavenumber of $k_w$.
Using the eigenfunctions of the whistler-mode proportional to $\exp i
(k_w x - \omega_0 t)$, the potential is rewritten as
\begin{equation}
\frac{\psi}{m_e c^2} \sim 
\frac{a_w^{+} a_w^{-}}{\widetilde{B}_{\rm ext} - 1} \;,
\end{equation}
where $a_w^{\pm} = e E_w^{\pm} / (m_e c \omega_0)$ is the normalized
electric field of the whistler waves. 
The external magnetic field considered in our analysis is typically
$\widetilde{B}_{\rm ext} \sim 10$--100 to avoid the electron cyclotron
resonance near $\widetilde{B}_{\rm ext} \sim 1$ \cite{sano17}.
Then the potential energy $\psi$ could be over 10 keV up to 1 MeV for the
relativistic amplitude of the whistler waves $a_w^{\pm} \gtrsim 1$. 
The electromotive potential is an attractive source for the energy
transfer from the waves to plasmas, which is a heart of the standing
whistler wave heating.

The ion temperature $T_i$ heated by this mechanism is described by
using the initial conditions, which are the target density 
$\widetilde{n}_{e0}$, the field strength $\widetilde{B}_{\rm ext}$,
and the laser amplitude $a_0$ \cite{sano19}. 
Supposing the accelerated ions are perfectly thermalized, the ion
temperature is given by  
\begin{equation}
\frac{k_B T_i}{m_e c^2} \sim 
\frac{2 \pi Z}3 \frac{a_0^2}
{(N + 1)^2 (\widetilde{B}_{\rm ext} - 1)} \;,
\label{eq:ti}
\end{equation}  
where $k_B$ is the Boltzmann constant, $Z$ is the ion charge.
Here the amplitude of the transmitted whistler wave is estimated by the relation of $a_w^{\pm} = 2 a_0 / (N + 1)$.
The refractive index of the whistler-mode is expressed as $N = [1 +
  \widetilde{n}_{e0}/(\widetilde{B}_{\rm ext} - 1)]^{1/2}$.
The index goes to unity in the strong field limit $\widetilde{B}_{\rm
  ext} \gg \widetilde{n}_{e0} > 1$, which means the target becomes
almost transparent for the CP lights.
In the high density limit $\widetilde{n}_{e0} \gg \widetilde{B}_{\rm ext} >
1$, the index is proportional to the square-root of the density, $N
\sim (\widetilde{n}_{e0} / \widetilde{B}_{\rm ext})^{1/2}$. 
Then the transmitted laser energy decreases as the target density
increases.
In any case, the strong magnetic field far beyond the critical
strength $B_c$ gives the higher whistler wave amplitude in the target.

To reach the saturation of ion acceleration through the positive
feedback loop, it requires a finite time
$\widetilde{\tau}_{\rm sat}$ \cite{sano19} that is evaluated by  
\begin{equation}
\widetilde{\tau}_{\rm sat} \sim
\left[ \frac{\pi}{16} \frac{m_i}{Z m_e}
  \frac{(N + 1)^2 (\widetilde{B}_{\rm ext} - 1)}
  {N^2 a_0^2} \right]^{1/2} \;.
\label{eq:tsat}
\end{equation}
Therefore the pulse duration $\tau_0$ should be longer than $\tau_{\rm sat}$ for the
efficient energy conversion.
The saturation timescale is determined by the ion motions so that the alternative expression below might be appropriate as it reflects
physical meaning,
\begin{equation}
  \omega_{pi} \tau_{\rm sat} \sim
  \left[ \frac{\pi}{16} \frac{(N + 1)^2 \widetilde{n}_{e0}
      (\widetilde{B}_{\rm ext} - 1)}
  {N^2 a_0^2} \right]^{1/2} \;,
\label{eq:tsat2}
\end{equation}
where $\omega_{pi}$ is the ion plasma frequency.

By contrast, the electron temperature caused by the resistive heating
is derived as 
\begin{equation}
\frac{k_B T_e}{m_e c^2} \sim
\left[
  \frac{40 \sqrt{2 \pi} \ln \Lambda }{9}
  \frac{Z r_e }{\lambda_0}
  \frac{\widetilde{n}_{e0} a_0^2}
{(N+1)^2 (\widetilde{B}_{\rm ext} -1)^2}
\widetilde{t}
\right]^{2/5} 
\label{eq:te}
\end{equation}  
\cite{leblanc14,sano19}, where $r_e = {e^2}/({4 \pi \varepsilon_0 m_e c^2})$
is the electron classical radius.
The electron temperature is also characterized by the initial
laser-plasma parameters $\widetilde{n}_{e0}$, $\widetilde{B}_{\rm
  ext}$, and $a_0$.
The ion temperature heated by collapsing standing whistler waves is
higher than the electron temperature in a wide range of parameters,
e.g., it is valid when $a_0 \gtrsim 0.3$ and $\widetilde{B}_{\rm ext}
\gtrsim 3$ for the case of $\widetilde{n}_{e0} = 50$. 

The longitudinal electric field drives the ion acceleration, so
that its amplitude is a key for efficient ion heating.
Theoretically, the electric field is given by
\begin{equation}
  a_x
  \sim -
  \frac{8 N a_0^2}{(N + 1)^2 (\widetilde{B}_{\rm ext} - 1)}
  \sin (2 k_w x) \;,
\label{eq:ex}
\end{equation}
so that ideally it increases in proportion to the square of the
laser amplitude $a_0$.
Here we consider the upper limit of the longitudinal electric field.
The electrostatic field $E_x$ appears due to the density fluctuations
of the electrons.
The maximum amplitude is evaluated by the Gauss's law
$\epsilon_0 \bm{\nabla} \cdot \bm{E} = e (Z n_i - n_e)$. 
Assuming the maximum electron condensation is expressed as $n_e =
n_{e0} [ 1 + \cos (2 k_w x) ]$ while the ion density is uniform $Z n_i
= n_{e0}$, then the maximum electric field is
obtained by $a_{\rm lim} \sim \widetilde{n}_{e0} / (2 N)$.
Furthermore, the longitudinal field cannot exceed that of the
original whistler wave $a_w$.
Thus the amplitude limit is written as
\begin{equation}
a_{\rm lim} \sim \min \biggl( \frac{\widetilde{n}_{e0}}{2
  N}, \frac{2 a_0}{N+1} \biggr) \;,
\end{equation}
which infers that the higher density is preferable to enhance
the upper limit of the longitudinal electric field and eventually to
increase the ion temperature.
In the high density limit, the upper limit is proportional to
$\widetilde{n}_{e0}/(2N) \propto (\widetilde{n}_{e0}
\widetilde{B}_{\rm ext})^{1/2}$. 
Therefore in order to increase the potential limit of $E_x$, all of the
key parameters ($\widetilde{n}_{e0}$, $\widetilde{B}_{\rm ext}$,
and $a_0$) are required to be as large as possible.

\section{Numerical Results \label{sec4}}

\subsection{A fiducial case in 2D}

\begin{figure}
\includegraphics[scale=0.6,clip]{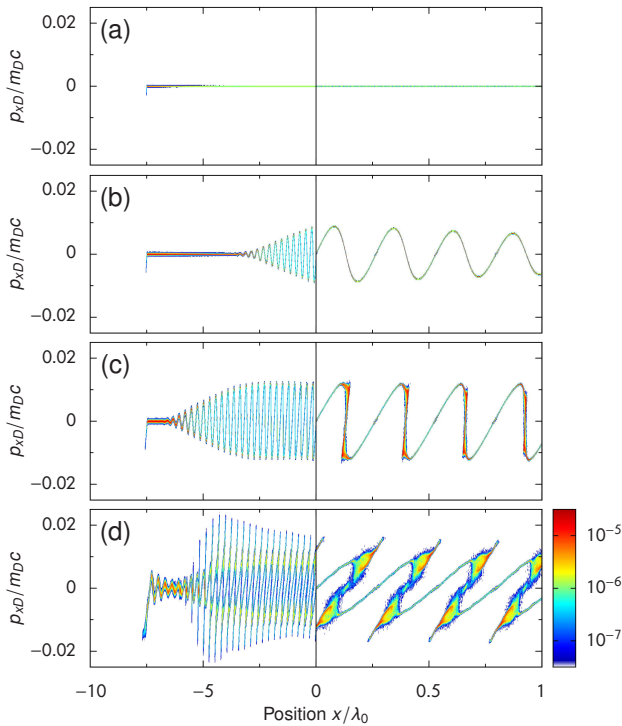}%
\caption{
Time evolution of the position-momentum phase diagram for the
deuterons during the formation and collapse of a standing whistler
wave in the 2D fiducial run.
The snapshot images are taken at (a) $\omega_0 t = 52$, (b) 83, (c)
114, and (d) 177.
The fiducial parameters are the DT target density $n_{e0}/n_c = 50$,
the external magnetic field $B_{\rm ext}/B_c = 20$, and the injected laser amplitude $a_0 = 4$.
The color denotes the particle number that is integrated with the $y$ direction.
The target plasma is located initially at $-7.5 \le x/\lambda_0 \le
7.5$.
The left panel shows the half region of the target at $x < 0$.
The right panel is the magnified view of the central part of the
target $0 \le x/\lambda_0 \le 1$. 
\label{fig2}}
\end{figure}

\begin{figure}
\includegraphics[scale=0.6,clip]{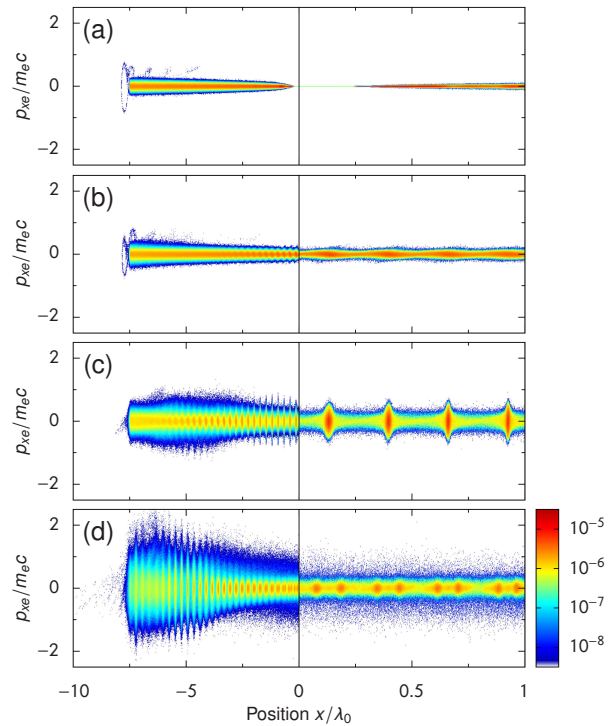}%
\caption{
Time evolution of the position-momentum phase diagram for the
electrons in the 2D fiducial run.
The model parameters and timing of the snapshots are the same as in
Fig.~\ref{fig2}.
\label{fig3}}
\end{figure}

Although the 1D picture clearly describes the essence of the standing whistler wave heating, it is important to
demonstrate whether it happens in more realistic multi-dimensional situations.  
Here we perform a typical case in 2D to observe the multi-dimensional effects. 

As the fiducial parameters, the initial density and thickness of a
DT target are selected as $\widetilde{n}_{e0} = 50$ and
$\widetilde{L}_x = 15$.  
A uniform external magnetic field is applied parallel to the laser
direction with the strength of $\widetilde{B}_{\rm ext} = 20$.
The laser intensity and pulse duration are $a_0
= 4$ and $\widetilde{\tau}_0 = 100$.  
The size of the computational domain is $(\widetilde{D}_x ,
\widetilde{D}_y) = (25, 3)$ in the $x$-$y$ plane.
For the fiducial parameters, the relation of the typical timescales is
$\omega_{pe} \gg \omega_{pi} \sim \omega_0$. 
The refractive index for the whistler wave is $N = 1.91$.
The amplitude of the transmitted whistler wave is $a_w = 2.75$ so that
large-amplitude waves move into the dense target.  
The pulse duration is longer than the growth
timescale of the ion velocity $\widetilde{\tau}_{\rm sat} \sim 45$
estimated from Eq.~(\ref{eq:tsat}) for the deuterons. 
Then the ion temperature is expected to increase to $T_i \sim
107$ keV by the standing whistler wave heating from Eq.~(\ref{eq:ti}).

The physical parameters in this model correspond to $n_{e0} = 5.57
\times 10^{28}$ m$^{-3}$, $L_x = 15$ $\mu$m, $B_{\rm ext} = 214$ kT,
$I_0 = 4.38 \times 10^{19}$ W/cm$^{2}$, $\tau_0 = 53$ fs by choosing
the wavelength $\lambda_0 = 1$ $\mu$m.
The target density is equivalent approximately to the hydrogen ice.
The laser conditions are within a range of standard femtosecond lasers
at present.
Notice that the outcome of the numerical simulations depends on the laser
wavelength even though the other non-dimensional parameters are
identical, which is because the Coulomb collision term has scale dependence.  

The 2D fiducial simulation reveals that the effects of multi-dimension and multi-ion species have little influence on the nature of the standing
whistler wave heating.
Figures~\ref{fig2} and \ref{fig3} show the position-momentum phase
diagram for the deuterons and electrons, respectively.
The deuteron distributions are depicted as the representative of the
ion species, because the behavior of the tritons is qualitatively
similar to that of the deuterons.
The quantities in these figures are integrated with the $y$ direction.
Surprisingly there is no dependence of the standing-wave evolution
in the $y$ direction.
The injected counter lasers arrive at the target surface when $t = 0$.
The snapshot data are taken at (a) $\widetilde{t} = 52$, (b) 83, (c)
114, and (d) 177.
The group velocity of the whistler waves is given by
\begin{equation}
  \widetilde{v}_g = \frac{2 N ( \widetilde{B}_{\rm ext} - 1 )^2}
  { 2 (
  \widetilde{B}_{\rm ext} - 1)^2 + \widetilde{n}_{e0} 
  \widetilde{B}_{\rm ext} } \;,
\end{equation}
which is $\widetilde{v}_g = 0.799$ for the fiducial parameters.
Then the two counter whistler waves meet at the center of the
target around $\widetilde{t} = 59$. 

The phase-diagram elucidates the continuous features of the ion-heating after the standing wave formation. 
Just before the standing wave sets in [Figs.~\ref{fig2}(a) and
  \ref{fig3}(a)], both of the whistler waves are traveling separately in the opposite direction. 
Until this stage, only the electrons are heated up by the resistive
heating due to the electron quiver motions of 
the whistler waves, while the ions remain quiet.
Drastic changes in the ion behavior begin at the moment when a
standing wave appears in the middle of the target.
A sinusoidal wavy structure appears in the ion phase diagram within
only five laser periods (about 15 fs).
The amplitude of the ion velocity increases linearly with
time, because the longitudinal field given by Eq.~(\ref{eq:ex}) is
independent of time.
The wavelength of the ion structure is exactly equal to the half of
the whistler wavelength $\widetilde{\lambda}_w / 2 = 0.262$.
The width of the standing wave expands toward the surface of the
target as the whistler waves propagate.
Then the ion acceleration region becomes more extensive, and the ions in the target gain more substantial energy in total.
The electron temperature increases gradually during the ion
acceleration stage associated with significant condensation of the electron
density [Figs.~\ref{fig2}(c) and \ref{fig3}(c)].
The saturation amplitude of the ion velocity is about 2\% of the speed
of light just before the ion-wave breakup.
Finally, the thermalization of ions starts through the ion two-stream
instability [Figs.~\ref{fig2}(d) and \ref{fig3}(d)].
The ion evolution in 2D is the same as the 1D cases presented
in the previous paper \cite{sano19}.

\subsection{Comparison between 2D and 1D results}

\begin{figure*}
\includegraphics[scale=0.9,clip]{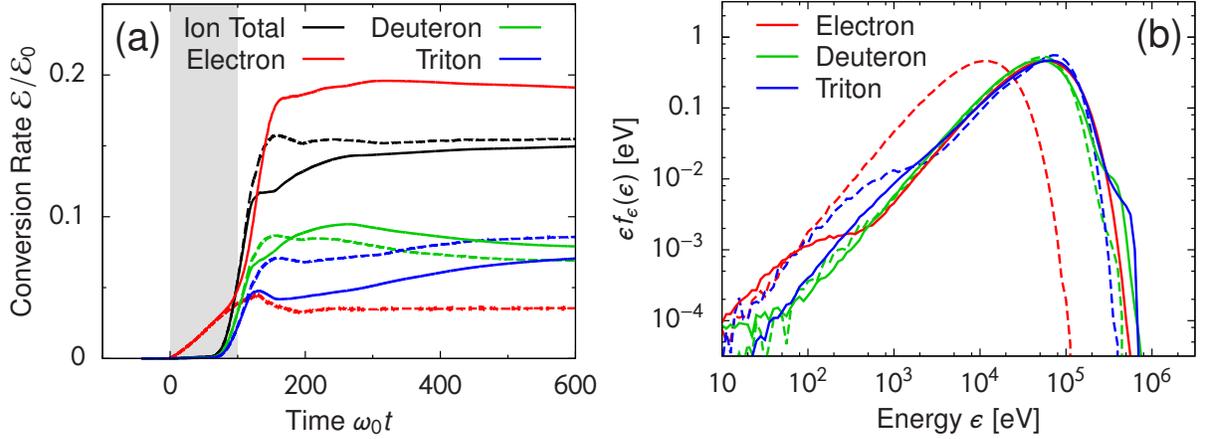}%
\caption{
(a)
Time histories of the plasma energies in the 2D fiducial run
(solid curves). 
The color indicates the different species of the target plasma, which are the electrons (red), deuterons (green), tritons (blue), and the total amount of all the ions (black).
For comparison, the results of the 1D simulation with the same model
parameters are also plotted by the dashed curves.  
The gray area stands for the laser pulse duration, where the injected
laser irradiates the target surface ($0 \le \omega_0 t \le 100$). 
(b)
Energy spectrum measured after the standing whistler wave
heating for the electrons (red), deuterons (green), and tritons (blue)
in the 2D fiducial run (solid curves).
The spectra are calculated at the end of the calculation $\omega_0 t = 1842$.
The dashed curves are the reference results of the 1D case with the
fiducial parameters.
\label{fig4}}
\end{figure*}

Here we clarify the similarities and differences between 2D and 1D
evolutions of the ion-heating mechanism focusing on the energy
conversion rate and the energy spectra.
We carry out a 1D simulation with the same initial conditions as in the
2D fiducial run.

The time profiles of the energy conversion in the 2D fiducial model
are shown by the solid curves in Fig.~\ref{fig4}(a).
The plasma energies are normalized by the laser energy ${\cal{E}}_0$
defined by the sum of the two lasers. 
As mentioned above, the electron energy starts to increase right after
the lasers hit the target at $t = 0$.
The rapid increase of the ion energies occurs lately initiated by the formation of the standing whistler wave.
The energy gain of the deuterons is slightly faster than that of
the tritons.
The origin of the difference is in the saturation time inferred by Eq.~(\ref{eq:tsat}), which has a dependence on the ion mass $m_i$.
The deuteron energy saturates earlier, and then the energy
increase of the tritons is disturbed by the turbulent motions of the
kinetic ion instabilities. 
However, during the thermalization phase, these ion energies tend to
relax into the isothermal equilibrium. 
The deuterons and tritons have similar energy eventually, and
the energy conversion rate to the total ions is settled in about 0.15.
The timescale of the ion thermalization takes a few times longer than
the saturation time $\widetilde{\tau}_{\rm sat} \sim 45$.

Figure~\ref{fig4}(a) shows the 1D results as well by the dashed curves
of which the same color denotes the same species. 
The qualitative behavior of ions is quite similar between the 2D and 1D cases.
The most significant difference is seen in the electron energy at the later stage.
In 2D, the electron energy is boosted up again together with the rapid increase of the ion energy.
The absorbed energy by the electrons is about 20\% of the injected
laser energy, which is higher than that of the total ions in the 2D case.
In contrast, the electron energy in the 1D case saturates when the laser pulse is finished, and the conversion rate is less than 5\%.
The enhancement of electron heating would be caused by the excitation of the electron plasma waves in the $y$ direction.

The energy spectra of the deuterons and tritons are unaffected by the
additional spatial dimension.
Figure~\ref{fig4}(b) shows the energy spectrum for each species at
$\widetilde{t} = 1842$, which is sufficiently after the thermalization.
The solid and dashed curves denote the 2D and 1D results, respectively.
In the 2D fiducial run, all three components exhibit a thermal 
distribution of nearly the same temperature.
The temperature of each component evaluated by the Maxwellian fitting
of the whole range is 32, 41, and 39 keV for the deuterons, tritons, and electrons.
These ion-temperatures agree roughly with the prediction by Eq.~(\ref{eq:ti}).
The simulation results clearly show that thermal fusion DT plasma is
generated by the counter irradiation of CP lasers under a strong
external magnetic field.

The ion temperatures in 1D, $T_D \approx 33$ keV and $T_T \approx 43$
keV, are almost identical to the 2D case.  
However, the electron temperature of $T_e \approx 7.8$ keV is much
lower than the 2D case. 
The electron temperature due to the resistive heating is $T_e \sim 6.24$
keV [Eq.~(\ref{eq:te})] so that it is consistent with the 1D result.
Therefore, it implies that some other effects besides the resistive
heating are contributed to the rise of the electron temperature in 2D. 

Based on the thorough comparison of various models, we conclude that
the ion-heating mechanism by the standing whistler waves can be
examined by 1D simulations adequately.  
The advantage of 1D analysis is that it allows detailed consideration
with more extended time integration and higher spatial resolution.
Hereafter we perform 1D PIC simulations to study parameter dependence
of the standing whistler wave heating.

\subsection{Optimization of the conversion efficiency}

The efficiency of the energy conversion from the whistler waves to
ions depends on the plasma and laser parameters. 
It is worth seeking the optimized conditions for the most efficient ion
heating.
For the fair comparison,
the DT target density $\widetilde{n}_{e0} = 50$,
the magnetic field strength $\widetilde{B}_{\rm ext} = 20$,
and
the laser amplitude $a_0 = 4$ are fixed
to ensure the predicted ion temperature is the same for all cases.
The critical parameters examined here are the target thickness
$\widetilde{L}_x$ and the pulse duration $\widetilde{\tau}_0$.

\begin{figure*}
\includegraphics[scale=0.9,clip]{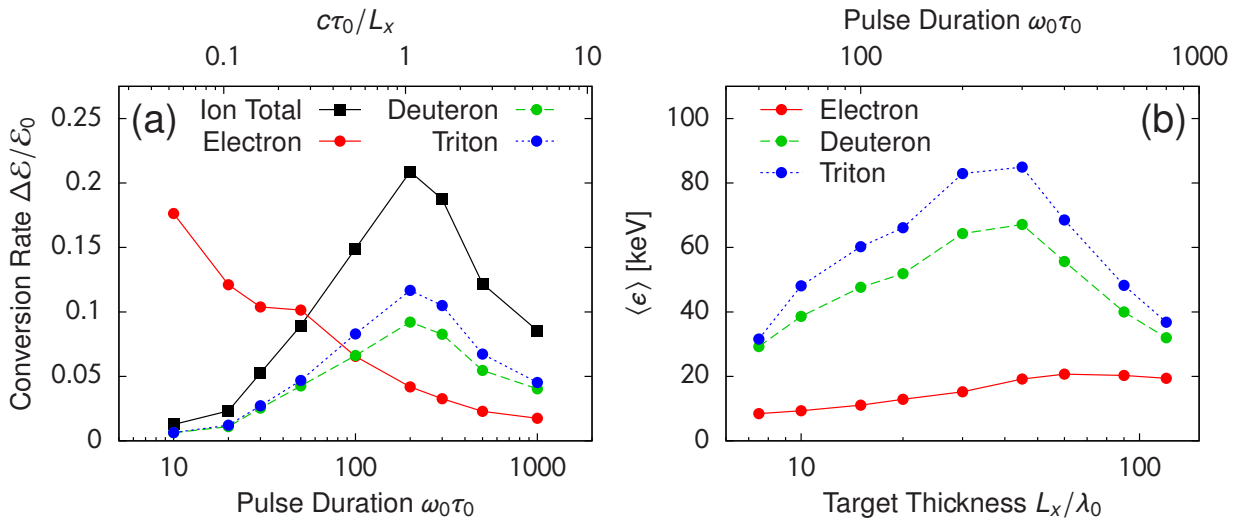}%
\caption{
(a)
Dependence of the laser pulse duration on the conversion efficiency
from the injected lasers to the target plasmas evaluated at $\omega_0
t = 2789$ by the 1D simulations.
The color indicates the different species which are the electrons
(red), deuterons (green), tritons (blue), and the total sum of the
ions (black).
The model parameters except for the pulse duration are identical for
all the runs.
The laser amplitude, field strength, and target density are the same
as in the 2D fiducial run, that is, ${n}_{e0}/n_c = 50$, ${B}_{\rm
  ext} / B_c = 20$, and $a_0 = 4$.
The target thickness adopted here is ${L}_x / \lambda_0 = 30$, which is
twice of the fiducial run. 
(b)
Dependence of the target thickness on the average energy of each
plasma species.
The ratio of the target thickness and laser pulse
duration is fixed to be $c \tau_0 / L_x = 10 / (3 \pi)$, or $\omega_0
\tau_0 = (20/3)(L_x/\lambda_0)$, in all the cases.
The model parameters are the same as those in the models shown by (a)
except for the target thickness and pulse duration.
\label{fig5}}
\end{figure*}

Figure~\ref{fig5}(a) shows the conversion efficiency to ions and
electrons as a function of the pulse duration $\widetilde{\tau}_0$ 
for the cases of $\widetilde{L}_x = 30$.
The plasma energies in this figure are measured at $\widetilde{t} =
2789$ from the 1D PIC simulations.
As can be seen, the efficiency for the total ions has a peak around
$\widetilde{\tau}_0 \sim 200$, which corresponds that the laser pulse
length is comparable to the target thickness $c \tau_0 \sim L_x$ or
$\widetilde{\tau}_0 \sim 2 \pi \widetilde{L}_x$.
Precisely, the pulse length of the whistler wave in the target should
be $v_g \tau_0$, but here we consider $v_g \sim c$ approximately. 
The conversion rates for the deuterons and tritons exhibit a similar
trend of the $\widetilde{\tau}_0$ dependence. 

In order for ions to gain the maximum energy, the standing whistler
wave should be sustained at least the saturation time
$\widetilde{\tau}_{\rm sat} \sim 45$. 
If the pulse duration is shorter than $\widetilde{\tau}_{\rm sat}$,
the ion heating must be inefficient because the ion acceleration is
restricted by the shorter lifetime of the standing wave.
The conversion efficiency increases until the pulse duration becomes a few times longer than $\widetilde{\tau}_{\rm sat}$.
If the pulse length is longer than the target thickness, the standing wave is formed in the vacuum region where there is no chance to
transfer the energy to the ions.
For those cases, the entire ions in the target are heated up to the
maximum temperature.
However, the wasted fraction of the laser energy becomes more
significant for the longer pulse duration.
Thus the efficiency of the energy conversion decreases as the pulse
duration increases.

As for the electron energy, the efficiency is decreasing monotonically 
with $\widetilde{\tau}_0$.
For the 1D cases, the electron temperature at ${t} = {\tau}_0$ given
by Eq.~(\ref{eq:te}) has a dependence of $\widetilde{\tau}_0^{2/5}$ so
that the efficiency is proportional to $\widetilde{\tau}_0^{-3/5}$,
which is roughly consistent with the simulation results.  

The best condition of the pulse duration in terms of the ion heating
will depend on the target thickness.
Next, the dependence of the ion energy gain on the target
thickness is examined.
We keep the ratio of $c \tau_0 / L_x$ constant, because
the highest efficiency is realized when $c \tau_0 \sim L_x$.
As seen from Fig.~\ref{fig5}(a), the most efficient case is
$\widetilde{\tau}_0 = 200$ for $\widetilde{L}_x = 30$, so that we use
this value as the reference ratio $\widetilde{\tau}_0 = (20/3)
\widetilde{L}_x$. . 
Figure~\ref{fig5}(b) shows the averaged kinetic energy for each
species $\langle \epsilon_{\ast} \rangle$, which is regarded as the temperature, as a function of
$\widetilde{L}_x$. 
The broad range of the parameters is considered from $\widetilde{L}_x = 7.5$
($\widetilde{\tau}_0 = 50$) to $\widetilde{L}_x = 800$
($\widetilde{\tau}_0 = 120$). 
The ion temperature is always much higher than the electron temperature.
While the electron temperature has a little dependence on the target
thickness, the ion temperature has a peak at around $\widetilde{L}_x
\approx$ 30--45. 

It is found that the duration time of the standing wave is a crucial factor in understanding the dependence on the target thickness.
If the standing wave exists shorter timescale than the saturation time
$\widetilde{\tau}_{\rm sat}$ of the ion acceleration, then the energy gain of the ions should be restricted.
The peak energy of $\langle \epsilon_T \rangle \approx 90$ keV is
consistent with the theoretical prediction given by Eq.~(\ref{eq:ti}).
The decline of the ion temperature at the larger $\widetilde{L}_x$ is caused by
the limitation in the penetration length of the whistler wave.

We perform a simple demonstration on the
whistler wave propagation to reveal the limit of penetration.
Prepare a sufficiently thick DT target of the fiducial density
$\widetilde{n}_{e0} = 50$ and inject a flat-top shape right-hand CP light with infinite pulse duration from one side of the target.
The external magnetic field with $\widetilde{B}_{\rm ext} = 20$ is 
applied in the laser-propagation direction.
In principle, the whistler waves penetrate overdense plasmas.
However, the wave propagation is prevented by the electron-ion
collisions and the stimulated Brillouin scattering (SBS), and then
the transmitted pulse duration has a finite limit. 

\begin{figure}
\includegraphics[scale=0.75,clip]{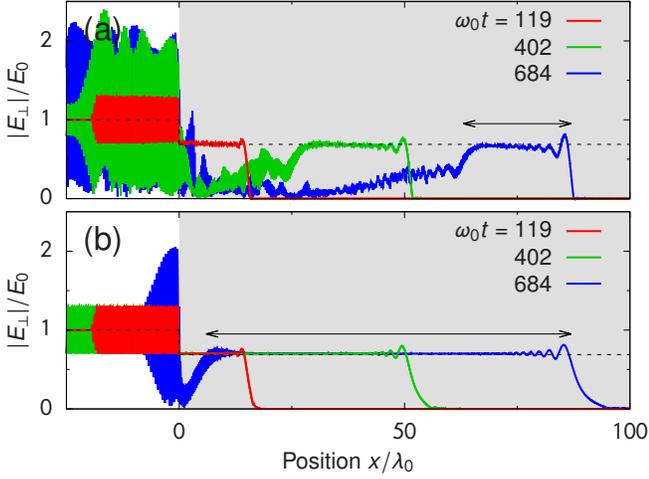}%
\caption{
Time evolution of the laser field penetrating an
overdense target for (a) collisional and (b) collisionless cases.
The solid curves show the envelope of the tangential electric
field of the laser.
The different colors indicate the different snapshot timings taken at
$\omega_0 t = 119$ (red), 402 (green), and 684 (blue).
The target thickness and pulse duration are assumed to be infinite.
The gray area denotes the target location at $x > 0$.
The flat-top shaped laser pulse reaches the target surface at $t = 0$.
The fiducial parameters (${n}_{e0}/n_c = 50$ , ${B}_{\rm ext}/B_c
= 20$, and $a_0 = 4$) are used.
The grid resolution is $\Delta x = \lambda_0 / 1000$, and the particle
number per grid is 100. 
The horizontal arrow stands for the pulse length of the whistler wave
entered in the target $L_w$.
\label{fig6}}
\end{figure}

Figure~\ref{fig6}(a) shows the tangential component of the electric field
of the injected laser $| E_{\perp} |$.
When the injected laser hits the target surface, a fraction of the wave energy gets into the target, and the rest is reflected.
The transmittance for the fiducial parameters is 0.688 which is shown
by the dashed line in the figure.
In the beginning, the whistler wave is propagating with the expected
amplitude in the target.
The position of the wavefront coincides with the group velocity of
the whistler wave $\widetilde{v}_g \approx 0.8$.
However, the penetration of the wave is stopped even
though the laser injection continues.
The pulse length of the whistler wave in the target is limited about
$\widetilde{L}_w \approx 25$, which corresponds to the pulse duration
of $\widetilde{\tau}_w = 2 \pi \widetilde{L}_w  / \widetilde{v}_g
\approx 200$. 
After ${t} \gtrsim {\tau}_w$, almost all the fraction of the injected wave is reflected at the target surface.
Then, a large-amplitude standing wave is formed outside of the target in the vacuum.
It is caused by the collisions between the electrons and ions.
Substituting the electron quiver motion ${v}_e = a_w c /
(\widetilde{B}_{\rm ext} - 1)$ and the electron density $n_e =
\widetilde{n}_{e0} n_c$
into the formula of the Coulomb collision frequency \cite{chen84}, 
\begin{equation}
  \nu_{ei} =
  \frac{\ln \Lambda}{4 \pi}
  \frac{Z e^4}{\varepsilon_0^2 m_e^2}
  \frac{n_{e}}{v_e^3}\;,
\end{equation}
the collision timescale $\tau_{\rm col} = \nu_{ei}^{-1}$ is derived as 
\begin{equation}
  \widetilde{\tau}_{\rm col} =
  \frac{4}{\pi \ln \Lambda}
  \frac{\lambda_0}{Z r_e}
  \frac{a_0^3}{(N+1)^3 \widetilde{n}_{e0}
    (\widetilde{B}_{\rm ext} - 1)^3}
  \; .
\end{equation}
For the fiducial parameters, the collision timescale becomes
$\widetilde{\tau}_{\rm col} = 344$, which is almost consistent
with the the duration time of the whistler wave in Fig.~\ref{fig6}(a).

There is another restriction on the transmitted wave duration, which is
the SBS of the whistler wave \cite{nishikawa68a,forslund72,lee74}.
In order to confirm the SBS effect, we perform another simulation in
which the collision term is switched off.
The collisionless model is shown in Fig.~\ref{fig6}(b).
All the other parameters are identical to those in Fig.~\ref{fig6}(a).
As expected, the whistler wave propagates much longer distances.
However, the transmittance is quenched suddenly by the growth of the
Brillouin instability.
The transmitted whistler wave decays into the back-scattered whistler
wave and the ion-acoustic wave.
The length and duration of the whistler wave are $\widetilde{L}_w
\approx 85$ and $\widetilde{\tau}_w
\approx 650$ in this case.  
The growth rate of the Brillouin instability is proportional to the
wave amplitude of the parent wave.
When we reduce the laser amplitude to $a_0 = 2$, the duration
of the whistler wave becomes $\widetilde{\tau}_w \gtrsim 800$.
On the other hand, the more intense laser makes the collision timescale longer.
The fine-tuning of the laser parameters would be required to maximize
the whistler wave duration.
Understanding the propagation properties of the whistler waves in
overdense plasmas would be a truly fundamental question.
However, it is beyond the scope of this paper, and we will address
the details of this topic in the subsequent paper.

\subsection{DT neutron yields}

In this subsection, the neutron yield is calculated theoretically
and numerically for the thermal fusion plasmas generated by the
standing whistler wave heating. 
Neutrons with the energy of 14.1 MeV are generated through the DT
fusion reaction,
\begin{equation}
\ce{D + T ->  \alpha \; (3.5 \; {\rm MeV}) + n \; (14.1 \; {\rm MeV})} \;.
\end{equation}
The DT plasma generated by the counter laser irradiation is an
attractive source of thermal neutrons because the ion temperature
becomes more than 10 keV.
It is found that the expected neutron yield could be comparable to or
even higher than the experimental results by the standard methods at 
present.   

The ion energy distribution is almost thermalized at $T_i \approx 40$ keV
after the standing whistler wave heating, so that the Maxwell-averaged
fusion reactivity is applicable in this circumstance \cite{atzeni04}.  
The reactivity of the DT reaction takes $\langle \sigma v \rangle_{DT}
\sim 8.0 \times 10^{-22}$ m$^{3}$/s at $T_i = 40$ keV. 
The neutron yield is evaluated by $Y_n = n_D n_T \langle \sigma v
\rangle_{DT} {\cal{V}} {\cal{T}}$, where ${\cal{V}}$ and ${\cal{T}}$
are the volume and lifetime of the fusion plasma.
Assuming the DT target density is $\widetilde{n}_e = 50$ for $\lambda_0 =
1$ $\mu$m, the number densities of the deuterons and tritons are given by
\begin{equation}
  n_D = n_T \sim 2.8 \times 10^{28}
  \lambda_{\mu{\rm m}}^{-2}
  \biggl( \frac{\widetilde{n}_{e0}}{50} \biggr)
  \;\; [{\rm m}^{-3}] \;,
\end{equation}
where $\lambda_{\mu{\rm m}}$ is the laser wavelength in the unit of
  $\mu$m. 
As for the plasma volume, it is estimated by a cylindrical volume
${\cal{V}} \sim \pi (d_0/2)^2 v_g \tau_0$ where $d_0$ is the diameter
of the laser spot and $v_g \tau_0$ is the length of the standing
whistler wave. 
The fusion reaction becomes
inactive gradually due to the decrease in the ion density associated
with the expansion. 
The timescale of the ion expansion would be equivalent to the propagation
time of the rarefaction wave inside of the target, that is, ${\cal T}
\sim v_g \tau_0 / (2 v_{\rm th})$ \cite{lindl95}.
In the direction perpendicular to the external magnetic field, the
expansion is suppressed by the ion gyro-motions and surrounded cold
and dense plasmas. 
The thermal velocity of deuterons is $\widetilde{v}_{\rm th} \sim 4.62 \times
10^{-3}$ for the temperature $T_D = 40$ keV.
The dimensional numbers of ${\cal{V}}$ and ${\cal{T}}$ are given by
\begin{equation}
  {\cal{V}} \sim 2.5 \times 10^{-17}
  \lambda_{\mu{\rm m}}^3
  \widetilde{v}_g
  \biggl( \frac{\widetilde{\tau}_0}{200} \biggr)
  {\widetilde{d}_0}^2
  \;\; [{\rm m}^3] \;,
\end{equation}
and
\begin{equation}
  {\cal{T}} \sim 11
  \lambda_{\mu{\rm m}}
  \widetilde{v}_g
  \biggl( \frac{\widetilde{\tau}_0}{200} \biggr)
  \;\; [{\rm ps}] \;.
  \label{eq:texp}
\end{equation}
Then the neutron yield is estimated as
\begin{equation}
  Y_n \sim 1.8 \times 10^{8}
  \widetilde{v}_g^2
    \biggl( \frac{\widetilde{n}_{e0}}{50} \biggr)^2
  \biggl( \frac{\widetilde{\tau}_0}{200} \biggr)^2
    {\widetilde{d}_0}^2 
  \;\; [{\rm n}] \;.
  \label{eq:yn}
\end{equation}

\begin{figure*}
\includegraphics[scale=0.9,clip]{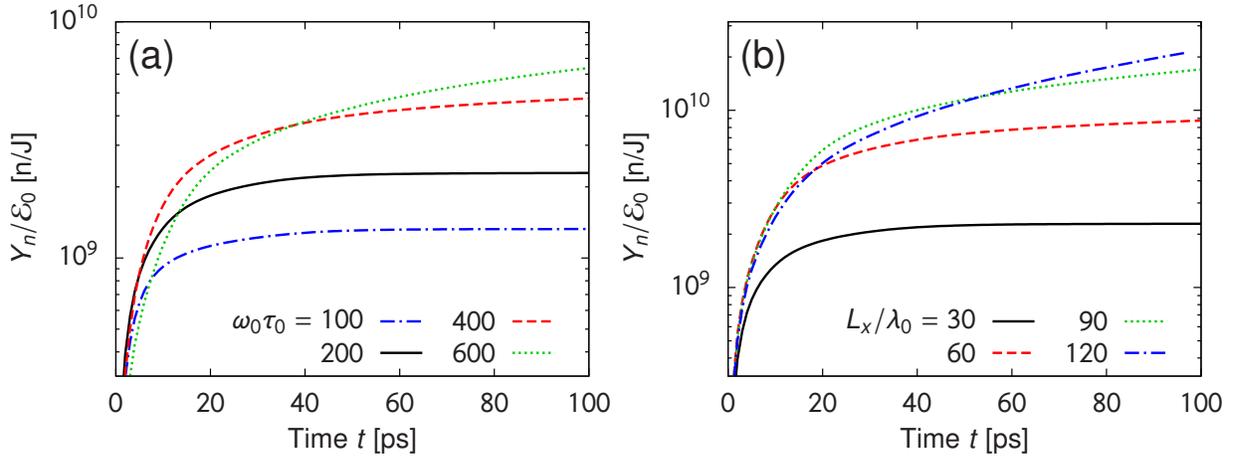}%
\caption{
(a)
Time profile of the DT neutron yield for various cases with
different pulse duration of $\omega_0 \tau_0 = 100$ (dot-dashed blue),
200 (solid black),
400 (dashed red), and 600 (dotted green).
The ratio of the target thickness and laser pulse
duration is fixed to be $L_x / \lambda_0 = (3/20) \omega_0 \tau_0$.
The neutron yield is normalized by the laser energy, so that it is
shown in the unit of n/J.
(b)
DT neutron yield for more efficient cases with different target
thickness of $L_x / \lambda_0 = 30$ (solid black), 60 (dashed red), 90
(dotted green), and
120 (dot-dashed blue). 
For these models, the same pulse duration $\omega_0 \tau_0 = 200$ is
used, which means that the difference in the neutron yeild is originated
from the different ratio of $c \tau_0 / L_x$.
\label{fig7}}
\end{figure*}

The laser-generated neutrons are often compared by using the neutron
generation efficiency, which is the neutron yield divided by the
injected laser energy, $\bar{Y}_n \equiv {Y_n}/{\cal{E}}_0$, where 
the laser energy ${\cal{E}}_0 = 2 \pi (d_0 / 2)^2 I_0 \tau_0$ for two
counter beams configuretion. 
The injected laser energy in the fiducial model becomes
\begin{equation}
  {\cal{E}}_0 = 7.3 \times 10^{-2}
  \lambda_{\mu{\rm m}}
  \biggl( \frac{a_0}{4} \biggr)^2
  \biggl( \frac{\widetilde{\tau}_0}{200} \biggr)
  {\widetilde{d}_0}^2
  \;\; [{\rm J}] \;.
\end{equation}
Finally, the neutron generation efficiency is obtained as
\begin{equation}
  \bar{Y}_n \sim 2.4 \times 10^9
  \lambda_{\mu{\rm m}}^{-1}
    \widetilde{v}_g^2
  \biggl( \frac{\widetilde{n}_{e0}}{50} \biggr)^2
  \biggl( \frac{a_0}4 \biggr)^{-2}
  \biggl( \frac{\widetilde{\tau}_0}{200} \biggr)
  \; \; [{\rm n}/{\rm J}] \;,
  \label{eq:yn2}
\end{equation}
which is independent of the laser spot size $d_0$.
The scaling of the efficiency suggests that the higher yield is
achieved when the density is higher or the duration of the whistler
wave is longer.
It should be noticed that the energy for generating the strong
magnetic field is not counted in the laser energy ${\cal{E}}_0$ by our
definition. 
Therefore, if the field generation over kT requires much more energy
than the heating laser, genuine efficiency may drop significantly
compared to the estimation by Eq.~(\ref{eq:yn2}).
The energy gain of $G$ is proportional to $\bar{Y}_n$.
For the fiducial case of DT ice target, the gain is quite small $G \sim 6.9
\times 10^{-3}$.
A feasible design for the fast ignition scheme of ICF by use of the
standing whistler wave heating is discussed later in the following
section. 

Next, the theoretical consideration of the neutron yield will be
tested by 1D long-term PIC simulations.
Figure~\ref{fig7} shows the time history of numerically
obtained neutron yield.
We consider the dependence of the neutron yield on the target
thickness and laser pulse duration.
The other parameters are
the same as in the fiducial run ($\widetilde{n}_{e0} = 50$,
$\widetilde{B}_{\rm ext} = 20$, and $a_0 = 4$).
The grid resolution of the models in
Fig.~\ref{fig7} is $\Delta x = 
\lambda_0 / 200$ and the initial particle number per grid is 100.
For the evaluation of the neutron yield, the computational domain is
divided into every 3 $\mu$m-thick bins.  
The average densities and temperature are calculated from the particles
in each bin, by which the yield per unit volume and unit time $n_D n_T
\langle \sigma v \rangle_{DT}$ is derived.
The neutron yields in the figure are the integrated quantities with the
volume and time. 

The neutron yields in the unit of n/J for four models picked up from
Fig.~\ref{fig5}(b) are shown in Fig.~\ref{fig7}(a).
The pulse duration for the models is $\widetilde{\tau}_0 = 100$, 200,
400, and 600.
The target thickness is determined from the optimized ratio of
$\widetilde{L}_x = (3 / 20) \widetilde{\tau}_0$.
When $\widetilde{\tau}_0 = 200$ ($\widetilde{L}_x = 30$), the yield is
about $2.3 \times 10^9$ n/J which is in good agreement with the estimation
given by Eq.~(\ref{eq:yn2}).
Theoretically, the normalized yield $\bar{Y}_n$ should be proportional
to $\widetilde{\tau}_0$ [Eq.~(\ref{eq:yn2})].
The numerical yield increases linearly from $1.3 \times 10^9$ to $4.8
\times 10^{9}$ n/J when the pulse duration is changed from
$\widetilde{\tau}_0 = 100$ to 400.
However it looks saturated for the case of $\widetilde{\tau}_0 = 600$.
It is caused by the propagation limit of the whistler wave
demonstrated in Fig.~\ref{fig6}(a). 
For the models in Fig.~\ref{fig7}(a), the entire region of the target
is heated up instantaneously.
The fusion plasma expands into the vacuum region and then the rapid
density decrease quenches the fusion reaction in the expansion
timescale of the order of ${\cal{T}} \sim 10$ ps.

The curious feature of the standing whistler wave heating is that
it occurs only at the standing wave.
If the pulse length is shorter than the target thickness, then the
standing wave region is smaller than the target thickness.
The heating will take place partly in the
middle of the target where the generated fusion plasma is surrounded
by the solid density with a cooler temperature.
Then the lifetime of the fusion plasma should be extended for the
thicker target cases, and the neutron yield will be enhanced
dramatically. 
Figure~\ref{fig7}(b) shows the time profile of $\bar{Y}_n$ for four cases
with different target thickness, $\widetilde{L}_x = 30$, 60, 90, and 120.
The laser pulse duration $\widetilde{\tau}_0 = 200$ is fixed for
all the models.
As seen from the figure, the neutron yield increases by a factor of
${\cal{F}} \sim 10$ as the target becomes thicker. 
Thus the ratio of the width of the standing wave and the target
thickness, $v_g \tau_0 / L_x$, affects the fusion plasma evolution.

\begin{figure*}
\includegraphics[scale=0.9,clip]{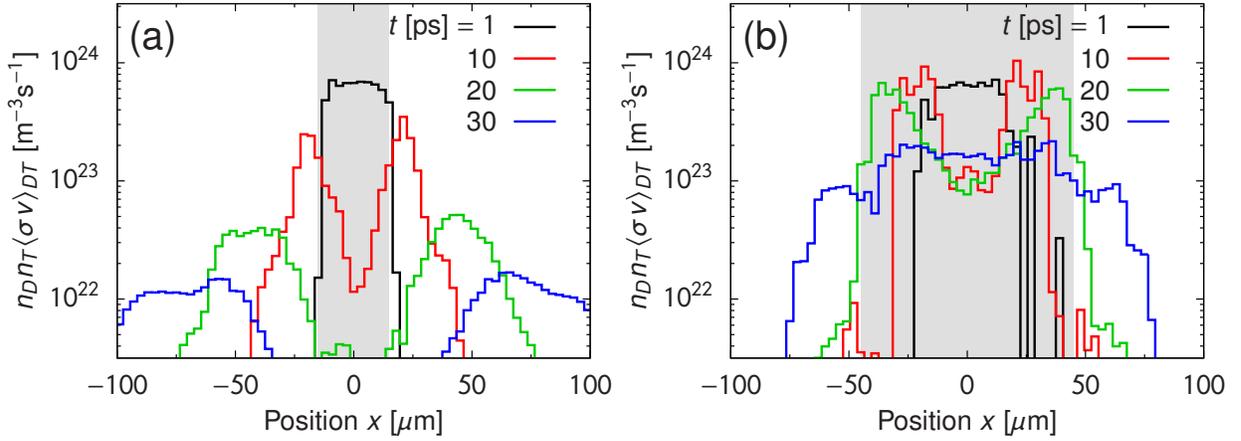}%
\caption{
Spatial distribution of the neutron generation rate per unit volume and
unit time for the cases of (a) $L_x / \lambda_0 = 30$ and (b) 90.
The snapshot data are taken at $t = 1$, 10, 20, and 30 ps after the laser
irradiation. 
The puluse duration is $\omega_0 \tau_0 = 200$ ($\tau_0 = 106$ fs) for
the both cases. 
The other laser-plasma parameters are the fiducial values such as
$n_{e0} / n_c = 50$, $B_{\rm ext} / B_c = 20$, and $a_0 = 4$.
The generation rate is evaluated from the average density and
temperature of the DT plasma in each 3-$\mu$m segments.
The temperature is estimated by two-third of the peak energy in the
energy spectrum $\epsilon f_{\epsilon} (\epsilon)$.
The gray area is the initial location of the DT target.
\label{fig8}}
\end{figure*}

Figure~\ref{fig8} shows the spatial distribution of the neutron
generation rate for the cases of $\widetilde{L}_x = 30$ and 90. 
When $\widetilde{L}_x = 30$ [Fig.~\ref{fig8}(a)], the pulse length is
equivalent to the target thickness, and then the ions in the entire
region are heated up at $t = 1$ ps.
The rarefaction waves reach to the target center at $t = 10$ ps, which
is consistent with the expansion timescale of ${\cal{T}}$ given by
Eq.~(\ref{eq:texp}). 
After that, the neutron generation rate decreases drastically due to
the decrease of the ion density. 
For the case of $\widetilde{L}_x = 90$ [Fig.~\ref{fig8}(b)], the high
temperature region at $t = 1$ ps is the central 30 $\mu$m of the target.
The neutron generation rate at the early phase is almost the same as
that in $\widetilde{L}_x = 30$ case. 
However, the hot region is surrounded by the cold and dense plasma, and then
shock-like structure travels inside of the target.
The high-temperature and high-density region exists far beyond the
simple expansion timescale ${\cal{T}}$.
The neutron generation rate is kept higher and the total neutron yield
is amplified by at least an order of magnitude.

The highest value of $\bar{Y}_n$ obtained in our numerical experiments is
$2.2 \times 10^{10}$ n/J for the case of ${L}_x = 120$ $\mu$m
($\widetilde{L}_x = 120$) and
${\tau}_0 = 106$ fs ($\widetilde{\tau}_0 = 200$).
{\color{black} Although an idealized simple setup is assumed in the
  simulations, the neutron generation rate by the standing whistler
  wave heating is higher than the current achievements of the laser
  experiments (see Appendix).} 
The most efficient case among the laser experiments is obtained from
an imploded thermal plasma by a mega-Joule laser \cite{lepape18}.
In general, the generation of thermal neutrons requires large-energy lasers more
than kilo-Joule class. 
The impact of our method is that it produces thermal fusion plasmas
with a typical size of femtosecond laser with less than 10 J,
only if a supercritical magnetic field is available.
This benefit may come to fruition as the most compact thermal neutron source with ultrahigh efficiency.

\section{Discussion \label{sec5}}

\begin{figure}
\includegraphics[scale=0.75,clip]{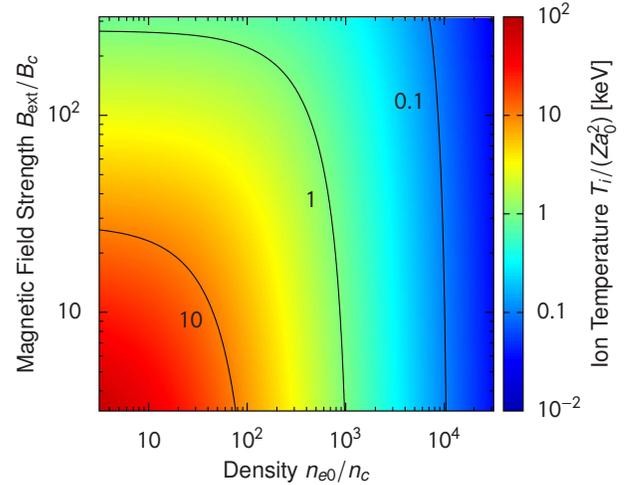}%
\caption{
Ion temperature predicted by Eq.~(\ref{eq:ti}) in the density and
magnetic field strength ($n_{e0}$-$B_{\rm ext}$) diagram. 
The color indicates the ion temperature of $T_i$ in the unit of keV,
which is divided by the non-dimensional laser intensity and ion
charge, $Z a_0^2$.
The contours are shown by solid courves at $T_i / (Z a_0^2) =
0.1$, 1, and 10.
\label{fig10}}
\end{figure}

\begin{figure*}
\includegraphics[scale=0.9,clip]{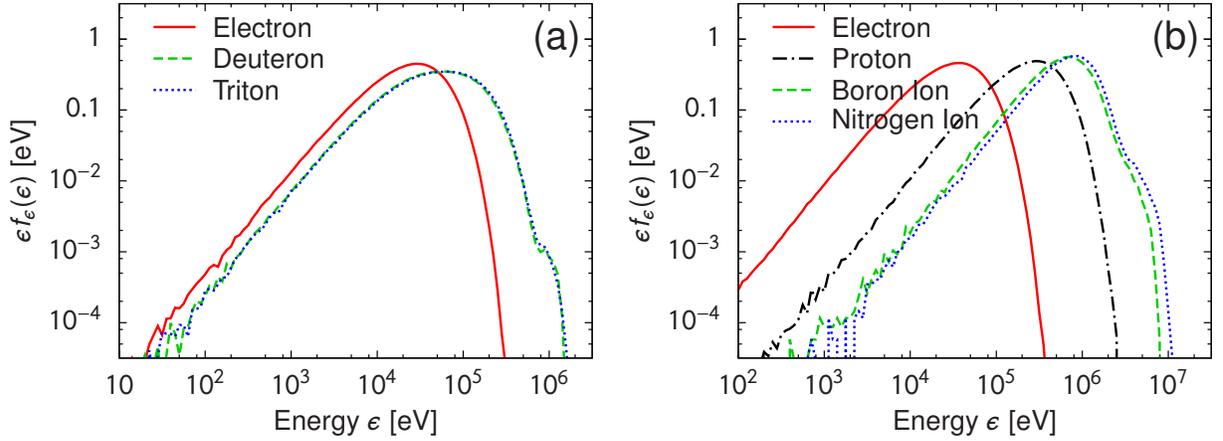}%
\caption{
(a)
Energy spectra in a case of imploded DT target heated by the
standing whistler wave for the electrons (solid red), deuterons
(dashed green), and
tritons (dotted blue). 
The key parameters of this run are the plasma
density $n_{e0}/n_c = 1500$, magnetic field strength $B_{\rm ext}/B_c
= 240$, and laser amplitude $a_0 = 20$.
The laser pulse duration is $\omega_0 \tau_0 = 200$.
The spectra are measured at $\omega_0 t = 1842$ after the laser
irradiation.
Taking account of the extremely high cyclotron frequency, the grid
resolution used in this run is $\Delta x = \lambda_0 /
2000$, and the particle number per grid is 100.
(b)
Energy spectra for a case of ammonia borane target.
Each species is shown by different colors, which are the electrons
(solid red), protons (dot-dashed black), boron ions (dashed green),
and nitrogen ions (dotted blue).
The simulation parameters of this run are the plasma density
$n_{e0}/n_c = 250$, magnetic field strength $B_{\rm ext}/B_c = 100$,
laser amplitude $a_0 = 12$, and pulse duration $\omega_0 \tau_0 =
200$.
The grid resolution is $\Delta x = \lambda_0 /
1000$ and the particle number per grid is 260.
\label{fig11}}
\end{figure*}

In this section, we will discuss the feasibility of the standing
whistler wave heating for the fast ignition scheme of ICF.
The counter configuration of heating lasers has been adopted in the
fast ignition scheme to increase the heating efficiency due to
self-generated magnetic fields induced by Weibel instability
\cite{kitagawa12,mori16,mori17}.
In our method, the counter beams are required to generate standing
waves in the fuels.
The inclusion of magnetic fields in the ICF scheme has been intensely
discussed \cite{hohenberger12,perkins13,wang15,fujioka16,sakata18,matsuo19},
because of the remarkable progress in the generation of laser-induced
magnetic fields \cite{yoneda12,fujioka13,korneev15,goyon17,santos18}. 
For example, a strong magnetic field may assist the ignition by 
stabilizing higher-mode Rayleigh-Tayler instabilities during the
implosion and suppressing the transverse heat conduction in hot spots
\cite{hohenberger12,perkins13}.
We also make use of magnetic fields for ion heating but in a different
concept.

For the fusion energy application, the energy gain is the most
important indicator. 
In order to increase the energy gain, the density
$\widetilde{n}_{e0}$ need to be higher or the pulse duration
$\widetilde{\tau}_0$ should be as long as possible.
Since the pulse duration of the whistler waves tend to have the limitation,
the density would be the only controllable parameter.
Figure~\ref{fig10} depicts the predicted ion temperature in the
diagram of the density and magnetic field strength.
The color denotes the ion temperature divided by the laser intensity
so that all the dependence on three key parameters ($\widetilde{n}_{e0}$,
$\widetilde{B}_{\rm ext}$, and $a_0$) is included in this figure.
In the high density limit, the ion temperature is proportional to
$T_i \propto a_0^2 / \widetilde{n}_{e0}$ and has little dependence on
$\widetilde{B}_{\rm ext}$.
Thus, even when the density is much higher than the DT ice density of
$\widetilde{n}_{e0} = 50$, the fusion plasma generation is possible
if the higher intensity laser is used for the ion heating.

{\color{black}
In the following, we consider the effective use of the standing
whistler waves in the heating part of the fast ignition scheme.
Notice that the implosion part is an assumption in our scenario so
that it might have some uncertainties and difficulties to be
confirmed.
However, it would be helpful as a guideline of the ICF research to
show the required quantities in the implosion for the attractive
application of our heating method.
}

\subsection{Possibility of an alternative ICF design}

A reasonable scenario for the enhancement of the energy gain
is the standing whistler wave heating of an imploded DT target by other
long-pulse lasers. 
In our setup, the existence of a uniform magnetic field is a necessary ingredient.
Then the cylindrical implosion perpendicular to the magnetic field line would be the best configuration.
For example, suppose the implosion of 30 times density enhancement.
Starting from the DT ice density, the imploded plasma condition is
$\widetilde{n}_{e0} = 1500$ ($n_{e0} = 1.67 \times 10^{30}$ m$^{-3}$).
We simulate the standing whistler wave heating of such an imploded DT core
plasma. 
Figure~\ref{fig11}(a) shows the energy spectrum of the DT core after
the counter irradiation of heating lasers.
The higher laser amplitude $a_0 = 20$ ($I_0 = 1.09 \times 10^{21}$
W/cm$^2$) than the fiducial case is necessary to keep the ion
temperature around 40 keV. 
For the higher energy conversion, the refractive index $N \sim
(\widetilde{n}_{e0} / \widetilde{B}_{\rm ext})^{1/2}$ should be
closer to unity, so that the stronger magnetic field is required.
We assume $\widetilde{B}_{\rm ext} = 200$ ($B_{\rm ext} = 2.14$ MT) in
this run.
The thickness of the core plasma and the pulse duration is assumed to
be $\widetilde{L}_x = 30$ ($L_x = 30$ $\mu$m) and $\widetilde{\tau}_0 = 200$
($\tau_0 = 106$ fs).
The key quantities in this run are $N = 2.92$,
$\widetilde{v}_g = 0.610$, and $\widetilde{\tau}_{\rm sat} \sim 25$.
The predicted temperatures are $T_i \sim 140$ keV for ions and $T_e
\sim 10.6$ keV for electrons.

The DT core is found to be heated up to 40 keV successfully.
The temperature of each species fitted by the Maxwellian distribution
is 43, 45, and 2.0 keV for the deuterons, tritons, and electrons. 
The energy gain of $G$ is estimated by the same assumption of the neutron generation efficiency $\bar{Y}_n$, which becomes
\begin{equation}
  G \approx 2.5
  \lambda_{\mu{\rm m}}^{-1}
    \widetilde{v}_g^2
  \biggl( \frac{\widetilde{n}_{e0}}{1500} \biggr)^2
  \biggl( \frac{a_0}{20} \biggr)^{-2}
  \biggl( \frac{\widetilde{\tau}_0}{200} \biggr)
    \biggl( \frac{{\cal{F}}}{10} \biggr)
  \;,
\label{eq:g}
\end{equation}
where we assume the amplification factor ${\cal{F}} \sim 10$ due to
the target thickness effect.
In the high density limit, the energy gain increases as $G
\propto \widetilde{n}_{e0} \widetilde{\tau}_0$ in keeping the ion
temperature $T_i \sim 40$ keV.
This is why we need to make the imploded density much higher or the
pulse duration of the whistler wave much longer to enhance the gain. 

The energy gain is managed to be higher than unity, although it
would be still insufficient as the power reactor.
Furthermore, only the heating lasers are counted in the estimation of
the injected laser energy.
The energy gain given by Eq.~(\ref{eq:g}) is valid only when the laser
energy for the heating is dominant compared to that for the implosion,
which will be required at least several kJ.
The implosion of magnetic-pressure dominant plasma is another serious difficulty of this scenario.
The magnetic field is also amplified through the compression by the same factor.
However, the magnetic pressure of the compressed 2 MT is of the order of
$10^{18}$ Pa, which is much higher than the thermal energy of 40 keV
plasma $\sim 10^{16}$ Pa.
It would be tough to find out a possible way to compress the cylindrical DT fuel against this enormous magnetic pressure.
In this sense, the self-generated magnetic field would be a more credible source of the mega-Tesla field formed during or after the implosion.
Empirically, the strength of the self-generated field is a few tens of
percent of the laser field, so that the generation of MT field
might be possible by high-intensity lasers of $a_0 \gtrsim 300$ ($I_0 \gtrsim 10^{23}$ W/cm$^2$) \cite{wang19}.
Anyhow, the order estimation of the energy gain [Eq.~(\ref{eq:g})]
encourages further deliberation in optimizing the design concept as an
alternative fast ignition by the standing whistler wave heating.

\subsection{Aneutronic proton-boron fusion reaction}

In modern fusion researches, the DT reactions are commonly considered because of the lowest reaction temperature.
However, the standing whistler wave heating achieves potentially much higher temperature plasmas, which bring a possibility of
ultraclean aneutronic fusion reaction such as the proton-boron
reaction, 
\begin{equation}
\ce{p + ^{11}B ->  3 \alpha \; (8.7 \; {\rm MeV})} \;.
\end{equation}
The thermal pB reaction works at the temperature over several hundreds of Kelvin.
The reactivity of pB reaction takes the maximum $\langle \sigma v
\rangle_{pB} \sim 3.9 \times 10^{-22}$ m$^{3}$/s at $T \sim 470$ keV
\cite{atzeni04,nevins00,putvinski19},
while the reactivity of DT reaction is the largest at $T = 67$ keV.
A few attempts of the pB fusion reaction have been reported by means
of laser-accelerated proton beams \cite{belyaev05,labaune13,picciotto14}.
Although the non-thermal ignition may reduce the requirement of
the ion temperature \cite{hora17}, we consider here the realization of
pB reaction initiated by the standing whistler wave heating with the
most straightforward consideration. 

For the pB fusion reaction, we can use the ammonia borane
\cite{demirci17} as the target material. 
The chemical formula of the ammonia borane is \ce{H_6BN}, which is
paid attention because of the higher hydrogen density than the
hydrogen ice and studied intensely as the hydrogen
storage material. 
The mass density is 780 kg/m$^3$, and the molecular weight is $A = 30.8$.
The number density of protons in ammonia borane is $n_p = 9.08 \times
10^{28}$ m$^{-3}$, whereas the number density of electrons is $n_e =
2.73 \times 10^{29}$ m$^{-3}$, which corresponds to $\widetilde{n}_e = 244$.
Another advantage as the target material is that it is in the solid-state at room temperature. 
Thus we examine whether the standing whistler wave could heat the
ammonia borane to be the thermal fusion plasma.

We perform a 1D PIC simulation expecting the pB reaction by
counter-laser irradiation in the same configuration as the DT case.
The target in the simulation is the ammonia borane plasma with the
electron density $n_{e0} = 2.79 \times 10^{29} $ m$^{-3}$
($\widetilde{n}_e = 250$) and the thickness is
$L_x = 30$ $\mu$m ($\widetilde{L}_x = 30$).
The laser intensity and pulse duration is assumed to be $I_0 = 3.94
\times 10^{20}$ W/cm$^2$ ($a_0 = 12$) and $\tau_0 = 106$ fs
($\widetilde{\tau}_0 = 200$).
The magnetic field is assumed to be $B_{\rm ext} = 1.07$ MT
($\widetilde{B}_{\rm ext} = 100$) in this model.
If the weaker magnetic field is adopted, the wave energy is transferred
to the electrons predominantly through the cyclotron resonance.
In this setup, the other important quantities are $N = 1.88$,
$\widetilde{v}_g = 0.825$, and $\widetilde{\tau}_{\rm sat} \sim 34$.

Figure~\ref{fig11}(b) is the energy spectra of the ammonia borane
target after the irradiation of counter lasers under the strong
magnetic field at $t = 978$ fs ($\widetilde{t} = 1842$).
Each species is thermalized with a different temperature.
The ion temperatures fitted by the Maxwellian distribution are $T_p \approx
190$, $T_B \approx 450$, and $T_N \approx 530$ keV for the protons, boron ions,
and nitrogen ions, while the electron temperature is $T_e \approx 2.4$
keV, which are consistent with the predicted temperatures of $T_p \sim
188$ keV for protons and $T_e \sim 7.70$ keV for electrons.
The collision timescale of 400 keV plasmas is longer than several
picoseconds.
The ion temperature heated by the standing whistler wave is
proportional to the ion charge $Z$ so that the numerical result of
the highest order of the ion temperatures is qualitatively consistent
with the theoretical model. 

The $\alpha$ particle is generated through this fusion reaction and
the yield is estimated by 
\begin{eqnarray}
Y_{\alpha} &=& 3 n_p n_B \langle \sigma v \rangle_{pB} {\cal{V}}
{\cal{T}} \\
&\sim&
1.1 \times 10^{10}
  \widetilde{v}_g^2
    \biggl( \frac{\widetilde{n}_{e0}}{250} \biggr)^2
    \biggl( \frac{\widetilde{\tau}_0}{200} \biggr)^2
    \biggl( \frac{\widetilde{d}_0}{10} \biggr)^2 \;,
\end{eqnarray}
Here we assume the ion temperature is $T_i = 400$ keV, the reactivity
is $\langle \sigma v \rangle_{pB} \sim 3.9 \times 10^{-22}$ m$^3$/s,
and the thermal velocity of the protons is $v_{\rm th} \sim 6.19
\times 10^6$ m/s ($\widetilde{v}_{\rm th} \sim 2.06 \times 10^{-2}$). 
{\color{black}
The cooling time by the Bremsstrahlung emission is
of the order of nanosecond \cite{rybicki79}, which is much longer than
the expansion time scale ${\cal{T}} \sim 2$ ps.}
The energy gain is quite low so that there must be many things to
do for the fusion energy application.
However, this process could be regarded as a compact $\alpha$ particle
generator, which will be useful for such as elemental analysis of
materials and the development of new materials. 
%


\section{Conclusion \label{sec6}}

We have investigated the generation of fusion plasmas by the standing
whistler wave heating using 2D and 1D PIC simulations.
The ion heating by collapsing standing wave is a unique wave-plasma
interaction that is capable of generating fusion plasmas quickly and
efficiently. 
The introduction of super-strong magnetic fields brings a new pathway of
the energy transfer from electromagnetic waves directly to ions in
overdense plasmas.
Our findings are summarized as follows.

\begin{enumerate}
\item
We confirm that the ion heating by the standing whistler wave works
for multi-ion species targets in multi-dimensional simulations. 
Thermal fusion plasmas are generated in a rather simple configuration
only if a static strong magnetic field over the critical strength
$B_c$ is available.
\item
The ion evolutions in 2D simulations are found to be almost identical
to those of the 1D cases.  
The efficiency of the direct energy conversion from the lasers to ions is
a few tens of percent.
The electrons are relaxed to the isothermal equilibrium with the ions
in the 2D situation.
\item
The ratio of the pulse duration and target thickness is the most crucial factor in optimizing the conversion efficiency.
When the pulse length $c \tau_0$ is comparable to the target thickness
$L_x$, the conversion efficiency from the injected lasers to ions
takes the maximum value.
\item
DT thermal fusion plasmas generated by our method are applicable to
compact and efficient neutron sources for various scientific and
industrial purposes. 
The neutron yield in the optimized model is expected to be over $10^9$
n/J, which exceeds the current experimental achievement of laser-generated
neutrons. 
\end{enumerate}

The standing whistler wave heating will provide an alternative
possibility for the fast ignition scheme of ICF if we can provide
strong magnetic fields.   
The detail fine-tuning of the target design utilized the standing
whistler wave heating would be an exciting subject to be pursued further.
The ammonia borane is found to be an idealistic target for the pB
thermonuclear fusion.
The protons and boron ions in the ammonia borane target are heated up
to about 500 keV by the standing whistler wave heating, where the aneutronic proton-boron reaction is highly expected.
{\color{black}
In this paper, 2D effects on the standing whistler wave heating are
examined in rather simple geometry. The extension to 3D consideration
brings curious questions such as the effect of a finite laser spot and
dependence on the incident angle of the colliding counter waves, which
are quite important issues for the future.
}

\begin{acknowledgments}
We thank Yuki Abe, Masayasu Hata, Natsumi Iwata, Yasuaki Kishimoto,
Ryosuke Kodama, Norimasa Ozaki,
and Youichi Sakawa for fruitful discussions. 
This work was partly performed under the joint research project of the
Institute of Laser Engineering, Osaka University. 
This work was supported by JSPS KAKENHI Grant No. JP15K21767,
No. JP16H02245, and No. JP19H01870 and by the JSPS Core-to-Core
Program, B. Asia-Africa Science Platforms.  
\end{acknowledgments}

\appendix*
\section{NEUTRON GENERATION EFFICIENCY}

\begin{figure}
\includegraphics[scale=0.9,clip]{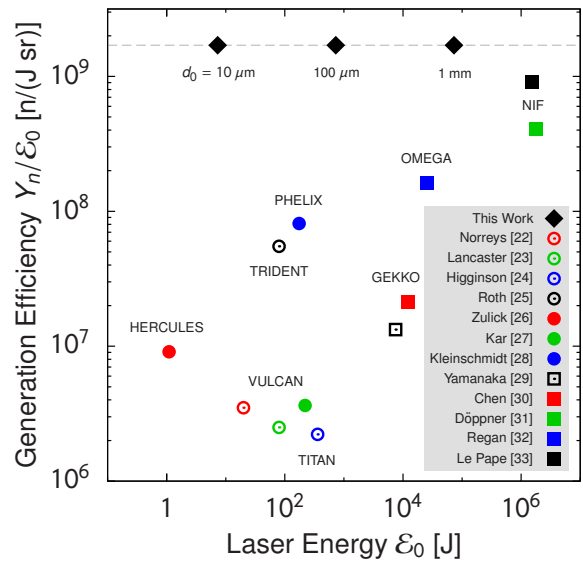}%
\caption{
Neutron generation efficiency of the unit of n/(J sr) in the optimized
model calculated from the 1D PIC simulation.
The highest neutron yield among our simulation results is 
$2.2 \times 10^{10}$ n/J which corresponds to $1.8 \times 10^9$ n/J
per steradian.
The 1D analysis has an ambiguity on the laser spot size so that three
cases are showed with the different assumption for $d_0 = 10$ $\mu$m,
100 $\mu$m, and 1 mm by the diamonds.
For comparison, the experimental achievements of laser-generated
neutrons done by various laser facilities are plotted in the figure.
The previous experiments are divided into two categories depending on
the method of neutron generation, which are the beam fusion method
(circles)
\cite{norreys98,lancaster04,higginson11,roth13,zulick13,kar16,kleinschmidt18}
and the implosion method (squares)
\cite{yamanaka86,chen90,doppner15,regan16,lepape18}.
The names of laser facilities used in the experiments are labeled near
the corresponding data points.
\label{fig9}}
\end{figure}

{\color{black}
The highest value of the neutron generation efficiency $\bar{Y}_n$
obtained in our numerical experiments 
is displayed in Fig.~\ref{fig9} together with the experimental data by
various laser facilities.
The solid angle averaged values of $\bar{Y}_n$ are shown in this figure.
The definition of the laser energy in our numerical models
includes only the heating counter lasers.
While the laser energy is proportional to the spot diameter $d_0$, the
normalized neutron yield $\bar{Y}_n$ is independent of $d_0$.
Therefore, the neutron yields obtained from the 1D PIC simulation can
be plotted at the different points depending on the choice of the
laser spot size $d_0$, which 
are $d_0$ = 10 $\mu$m, 100 $\mu$m, and 1 mm from left to right in the figure.
The experimental data include the results by the beam fusion method,
which is so-called the pitcher-catcher method
\cite{norreys98,lancaster04,higginson11,roth13,zulick13,kar16,kleinschmidt18}
and the implosion method
\cite{yamanaka86,chen90,doppner15,regan16,lepape18}.
}


%


\end{document}